\documentclass[prc,amsmath,amssymb]{revtex4}
\usepackage{epsfig,graphicx,cancel,bm}
\usepackage{tikz}
\usepackage{color}
\usepackage{verbatim}
\usetikzlibrary{matrix,arrows}
\begin{document}
\title{Parity nonconservation in deuteron photoreactions}
\author{T.~M.~Partanen}
\email{tero.partanen@helsinki.fi}
\author{J.~A.~Niskanen}
\email{jouni.niskanen@helsinki.fi}
\affiliation{
Department of Physical Sciences, P.~O.~Box~64, FIN-00014
University of Helsinki, Finland}
%
%\date{\today}
%
\begin{abstract}
We calculate the asymmetries in parity nonconserving deuteron 
photodisintegration due to circularly polarized photons
$\vec{\gamma}d\rightarrow np$ with the photon laboratory 
energy ranging from the threshold up to 10 MeV and the radiative
capture of thermal polarized neutrons by protons 
$\vec{n}p\rightarrow\gamma d$. We use the leading order 
electromagnetic Hamiltonian neglecting the smaller nuclear 
exchange currents.
Comparative calculations are done by using the Reid93 and Argonne
$v_{18}$ potentials for the strong interaction and the DDH and 
FCDH "best" values for the weak couplings in a weak one-meson 
exchange potential. A weak $N\Delta$ transition potential is 
used to incorporate also the $\Delta(1232)$-isobar excitation in 
the coupled-channels formalism.  
\end{abstract}
\maketitle
\section{\label{sec1}Introduction}
Due to the incessant presence of the strong interaction, the
hadronic weak interaction in the quark flavour conserving sector
of the Standard Model is not completely understood. Even though
the weak interaction between quarks is well known at high
energies, its properties are hard to extract in the nonperturbative
regime of quantum chromodynamics (QCD). This is due to the complicated
structure of nuclear systems along with coinciding dynamics of QCD.
It seems an almost hopeless challenge to distinguish directly
about the size of seven orders of magnitude smaller weak interaction
effects from those of QCD. Fortunately, the weak interaction leaves
a unique signature in the form of parity nonconservation (PNC),
which provides a tiny but non-vanishing observable.

For the past three decades the PNC calculations between nucleons
have been based on the use of the DDH \cite{ddh} potential. Unfortunately,
the potential requires the knowledge of several weak meson-nucleon
coupling constants, which are still to date inadequately known.
Today's contemporary attempts to determine PNC amplitudes
more and more often harness QCD based chiral perturbation theory
($\chi$PT). Unlike the phenomenological meson-exchange
model, $\chi$PT provides a systematic and model-independent way
to study hadronic reactions at low energies.
The updated $\chi$PT based approach leads to a systematic
expansion of PNC amplitudes in terms of low-energy constants
(LECs) which have a straightforward correspondence with the
weak DDH meson-nucleon couplings, see {\it e.g.} refs. \cite{zhu,ram}.

Out of all the possible candidates for asymmetry
and polarization observables,
in this paper we focus on the asymmetries associated with PNC
deuteron photodisintegration by circularly polarized photons
$\vec{\gamma}d\rightarrow np$ and thermal polarized neutron capture
$\vec{n}p\rightarrow\gamma d$. It is further to be noted that the
photon polarization of $np\rightarrow\vec{\gamma}d$ at threshold
equals the photon asymmetry of the time-reversed reaction 
$\vec{\gamma}d\rightarrow np$ at threshold.
The threshold behaviour of the asymmetry/polarization in reactions
$\vec{\gamma}d\leftrightarrow np$ can be shown to be insensitive to
the $\pi$-meson exchange, which represents the long-range $\Delta I=1$
part of the PNC interaction.
Therefore, the threshold region is essentially dominated by the
exchanges of heavy mesons ($\rho$, $\omega$) and thus also
the relatively long-ranged
$\Delta$-excitation could occur more clearly highlighted than what
it would if it appeared in a background where the pion is more
intensely present. To our knowledge this effect has only been checked
in the form of exchange currents in refs.\cite{schiacar,schia}.
Contrary to the reactions $\vec{\gamma}d\leftrightarrow np$
at the threshold, the PNC $\pi$-exchange is predominant in the
low-energy reaction
$\vec{n}p\rightarrow\gamma d$.

There are various theoretical works on the PNC reactions
$\vec{\gamma}d\rightarrow np$ 
\cite{lee,oka,khrip,fuji,liu,schia,hyun,liu07},
$np\rightarrow\vec{\gamma}d$ 
\cite{desp,craver,haid,hyun2,schindler,ando,liu07,lasmckell},
and
$\vec{n}p\rightarrow\gamma d$
\cite{tadic,danilov,lassey,despnp,schindler,kaplan,savage,park,
schia,haid,ando,hyandes,hyun2pi,liu07,schiacar}.
The calculations are typically carried out exploiting the old
meson-exchange picture, aside from some of the recent works, 
which apply the modern state-of-art
techniques, {\it e.g.} such as the
pionless effective field theory EFT($\cancel{\pi}$) and heavy-baryon
chiral perturbation theory HB$\chi$PT. The
results more or less agree
in the threshold region.
The energy regime of several MeV above the threshold of the
deuteron photodisintegration is investigated in refs.
\cite{oka,khrip,fuji,liu,schia}. Again, in that regime, the
results are similar except in ref. \cite{oka}, which differs by an 
exceptionally large pion contribution. 
The difference is discussed in detail in ref. \cite{liu}.
Up to date, for the reaction $\vec{\gamma}d\rightarrow np$, there
exist only two experimental data points: $(7.7\pm5.3)\times10^{-6}$
and $(2.7\pm2.8)\times10^{-6}$ at the photon energies of 3.2 and
4.1 MeV respectively \cite{earle}.
The latest photon polarization measurement for the inverse reaction
$np\rightarrow\vec{\gamma}d$ gives the value of $(1.8\pm1.8)\times10^{-7}$ \cite{knyaz}.
The data from the 1980's for both the reactions are
consistent with zero with rather a poor precision and,
therefore, new and more accurate experimental data,
{\it e.g.} \cite{jlab,stiliaris}, would be appreciated.

The asymmetry of the reaction $\vec{n}p\rightarrow\gamma d$ was
also measured previously in the 70's and 80's, resulting in the 
insufficiently accurate values of
$(0.6\pm2.1)\times10^{-7}$ \cite{cavai} and
$-(1.5\pm4.8)\times10^{-8}$ \cite{alberi}.
However, there is currently an ongoing experiment (NPDGamma) on the 
reaction $\vec{n}p\rightarrow\gamma d$ with cold neutrons at LANSCE 
and SNS with a preliminary result, which sets the asymmetry within the limits of $(-1.2\pm2.1({\rm stat.})\pm0.1({\rm sys.}))\times10^{-7}$
\cite{npdgamma}.
The oncoming measurements of the NPDGamma experiment aim to
improve the accuracy up to a level of 20\% of the typical theoretical
prediction $-5 \times 10^{-8}$, which
employs the DDH "best value" for the weak $\pi NN$ coupling constant 
$h^{(1)}_\pi$.
The asymmetry is straightforwardly proportional to the $h^{(1)}_\pi$ 
and therefore the current experiment is about to shed some light on 
the uncertain value of the coupling.

In this paper we study PNC in the above reactions using two
modern phenomenological strong potentials and post-DDH weak
couplings. Furthermore, we want to estimate the size of the
aforementioned $\Delta$ effect, which was found to be significant
in PNC elastic $\vec{p}p$ scattering at higher energies 
\cite{iqbalpnc,nispnc}. As in these works our calculation is 
carried out within the framework of the coupled channels 
meson-exchange model and hence, according to the common practice, 
we utilize the DDH potential \cite{ddh} as the starting point for 
the PNC $\pi$-, $\omega$-, and $\rho$-exchanges extending to use 
the weak couplings of ref. \cite{fcdh} consistent with the presence
of the $\Delta$.
We take account of exchange currents only in the extent they
exist when the Siegert's theorem \cite{siegert} is applied. They
have been considered more explicitly {\it e.g.} in refs. 
\cite{liu,schiacar,schia}.
In practise we use the electromagnetic Hamiltonian under the 
dipole approximation, which allows the PNC deuteron breakup to 
have four $pn$ continuum channels ${}^1S_0, {}^3P_0,{}^3P_1,$ 
and ${}^3P_2-{}^3F_2$ in the expansion up to $P$ waves.

This paper is organized as follows. In sect. \ref{sec2},
we give the appropriate forms of the electromagnetic Hamiltonian,
the  scattering and deuteron wavefunctions, and the PNC one-meson
exchange $NN$ and $N\Delta$-transition potentials.
The forms of the spin observables
$\mathcal{A}_{\vec{\gamma}}$ and $\mathcal{A}_{\vec{n}}$
respectively of the reactions $\vec{\gamma}d\rightarrow np$ and 
$\vec{n}p\rightarrow\gamma d$ are also given.
Summary of the results and conclusions are presented in sect.
\ref{sec3}.
\section{\label{sec2}Theory}
\subsection{\label{subsec2}Interactions}
\begin{figure}[t]
\includegraphics[width=16.cm]{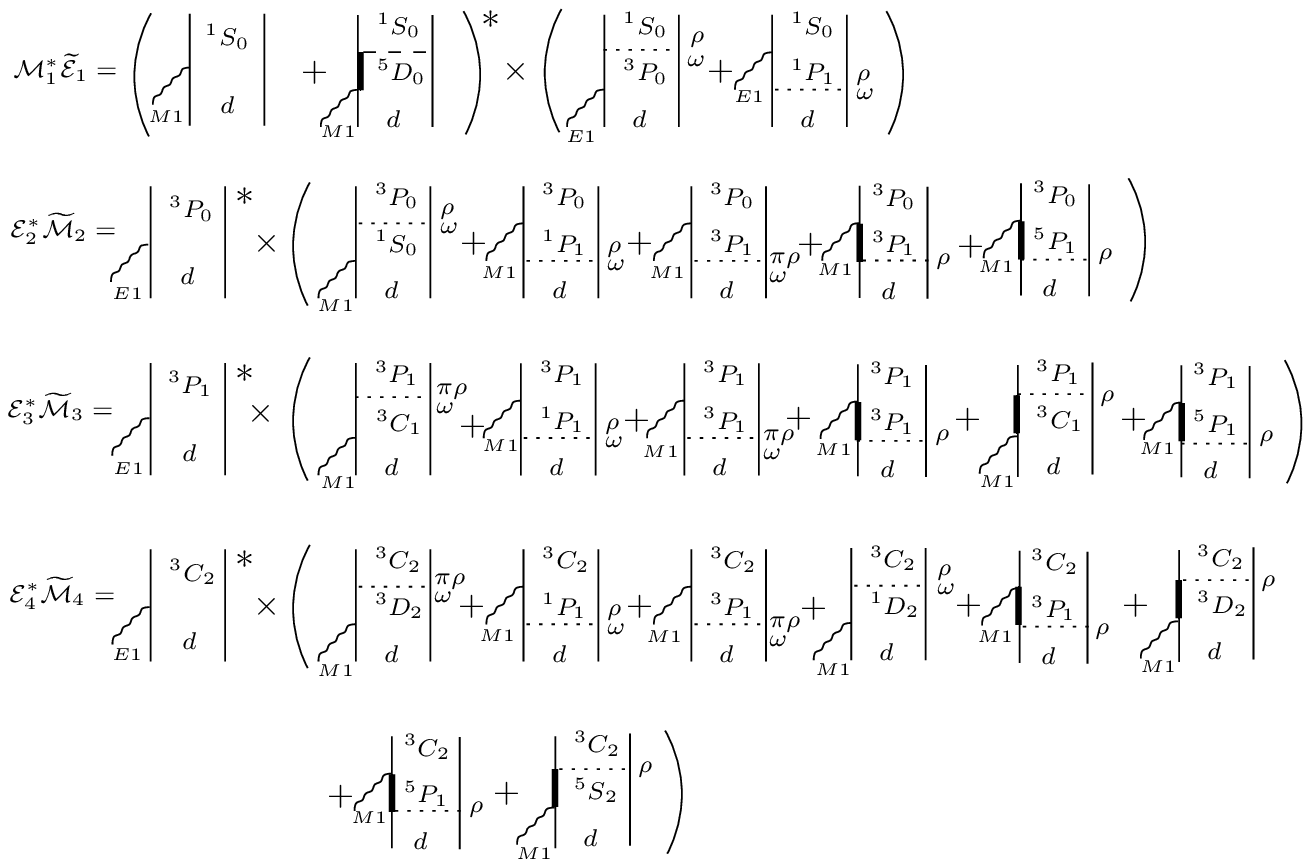}
\caption{\label{graphs}
Graphical representation for the $np$ and the direct
$N\Delta$ contributions in the deuteron photodisintegration.
The dashed line represents the strong interaction, the
dotted line the weak interaction, and the wavy line the
incoming photon. For brevity the tensor coupled scattering states ${}^3S_1-{}^3D_1$
and ${}^3P_2-{}^3F_2$ are denoted by ${}^3C_1$ and ${}^3C_2$
respectively. Similarly, the $d$ stands for the standard PC deuteron
${}^3S_1-{}^3D_1$ state.}
\end{figure}

The observable asymmetries in the photoreactions 
arise from an interference between the strong and weak
interactions giving rise to simultaneous presence of the
photomagnetic and photoelectric effects.
The interference appears when these two parallel processes
share the same final state quantum numbers,
and the PNC observables are obtained in terms of products
of PC and PNC partial wave amplitudes. This is illustrated in
fig. \ref{graphs}, where the mechanisms ($NN$, $N\Delta$
with PC and PNC forces) are shown alongside with the
quantum numbers of the possible states up to $P$-wave final
nucleons. The electromagnetic perturbing Hamiltonian
$\hat{H}_{\rm e.m.}=\hat{H}_{\rm E1}^{NN}+
\hat{H}_{\rm M1}^{NN}+\hat{H}_{\rm M1}^{N\Delta}$ which takes
care of the disintegration and formation of the deuteron is 
considered in the dipole approximation 
$e^{\pm i\bm{k}_\gamma\cdot\bm{r}}\approx1$.

The Hamiltonian for absorption of a photon may be written as
\begin{equation}\label{emham}
\hat{H}_{\rm e.m.}^\lambda=
-i\sqrt{\frac{\alpha\pi\omega_\gamma}{2}}
\hat{\bm{\epsilon}}_{\bm{k}_\gamma\lambda}\cdot\sum_i
\Bigl[(1+\hat{\tau}_i^z)\bm{r}_i
+\frac{1}{2M}\Bigl((1+\hat{\tau}^z_i)\bm{\ell}_i
+(\mu_s+\mu_v\hat{\tau}^z_i)\bm{\sigma}_i
+2\mu^\star(\hat{T}_i^z\bm{S}_i+{\rm h.c.})\Bigr)
\times\hat{\bm{k}}_\gamma
\Bigr],
\end{equation}
where $\hat{\bm{\epsilon}}_{\bm{k}_\gamma\lambda}$ ($\lambda=\pm1$)
is the circular polarization vector of the incoming photon,
$\omega_\gamma$ the center-of-mass energy of the photon, $\mu_s=0.88$
and $\mu_v=4.71$ the isoscalar and isovector magnetic moments
of nucleons, $\alpha=e^2$ the fine-structure constant, and $M=939$
MeV the average nucleon mass.
The $\gamma N \Delta$-vertex can occur in the presence of the
M1 and E2 transitions. We consider only the dominant M1
multipole and neglect the small E2 effect. The value of the
transition magnetic moment is given by the quark model as
$\mu^\star=f^\star\mu_v/2f=3\sqrt{2}\mu_v/5$ \cite{ericwei}.
The nucleon-Delta spin and isospin transition operators are
denoted as $\bm{S}$ and $\hat{T}_z$ \cite{brown}.

In the presence of possible channel coupling
the final state scattering wavefunctions are of the form
\begin{equation}\label{scatwf}
\psi_{SM_S}^{T(-)}(\bm{k},\bm{r})=
\frac{4\pi\sqrt{2}}{kr}
\sum_{\kappa'JM}\sum_{LM_L}
i^{L}\langle LM_LSM_S\vert JM\rangle
Y^{\ast}_{LM_L}(\hat{\bm{k}})
\mathcal{U}_{\kappa\kappa'}^{J(-)}(k,r)
\mathcal{Y}^{L'S'}_{JM}(\hat{\bm{r}})\vert T'0\rangle,
\end{equation}
where the superscript $"(-)"$ on the wavefunctions refers to
the incoming wave boundary conditions. The $\vert T0\rangle$
are the relevant isospin states for the $pn$-interaction,
$\mathcal{Y}^{LS}_{JM}(\hat{\bm{r}})$ the eigenfunctions of the
coupled total angular momentum, and
$\mathcal{U}^{J(-)}_{\kappa\kappa'}(k,r)$ the complex-valued
radial wavefunctions.
The quantum numbers $LST$ (relative orbital
angular momentum $L$, total spin $S$, and total isospin $T$), which
may be changed by the nuclear forces,
are denoted for brevity by $\kappa$ which also refers to the main 
final wavefunction, whereas $\kappa'$ labels the "small" component
generated from it.
The $J$ and $M$ relate to the total angular momentum
and they are good quantum numbers under the nuclear forces.
The calculated matrix elements 
${}^{(-)}\langle\bm{k};SM_ST|
\hat{H}_{\rm e.m.}^\lambda|M_d\rangle_\mathcal{D}$
will be expressed in terms of the
wavefunctions with the outgoing boundary conditions $"(+)"$,
{\it i.e.} $\mathcal{U}^{J(+)}_{\kappa\kappa'}(k,r)
=\mathcal{U}^{J(-)\ast}_{\kappa\kappa'}(k,r)$.

The deuteron wavefunction consists of a superposition of six
relevant components:
the standard PC $NN$ (${}^3S_1$, ${}^3D_1$)
and the tiny PNC $NN$ (${}^1P_1$, ${}^3P_1$)
and $N\Delta$ (${}^3P_1$, ${}^5P_1$)
states and may be written as
\begin{equation}\label{bound}
\psi_{M_d}^d(\bm{r})=
\sum_{\kappa_d}\frac{\mathcal{D}_{\kappa_d}(r)}{r}
\mathcal{Y}_{1M_d}^{L_dS_d}(\hat{\bm{r}})\vert T_d0\rangle,
\end{equation}
with the normalization $\int_0^\infty
dr\sum_i^6|\mathcal{D}_i|^2=1$.
The PC bound wavefunctions are real-valued and
their PNC partners imaginary-valued.
Both the scattering and bound state wavefunctions are obtained
by solving coupled Schr\"{o}dinger equations. The radial
Schr\"{o}dinger equation for the parity-admixed interaction
reads
\begin{align}\label{schreq}
\Bigl(\frac{\partial^2}{\partial r^2}
&-\frac{L'(L'+1)}{r^2}
+k^2\Bigr)\mathcal{U}^{J(+)}_{\kappa\kappa'}(k,r)\nonumber\\
&=2\mu\sum_{\kappa''}\int d\Omega_{\bm{r}}
\mathcal{Y}^{JM\dagger}_{L''S''}(\hat{\bm{r}})
\langle T''0|\hat{V}^{{\rm PC}}({\bm{r}})
+\hat{V}^{{\rm PNC}}({\bm{r}})|T'0\rangle
\mathcal{Y}^{JM}_{L'S'}(\hat{\bm{r}})
\mathcal{U}^{J(+)}_{\kappa\kappa''}(k,r),
\end{align} 
where for $\hat{V}^{{\rm PC}}({\bm{r}})$ the digonal $NN$ potential 
is taken as the phenomenological updated Reid soft core (Reid93) 
\cite{reid93} and Argonne $v_{18}$ (A$v_{18}$) \cite{av18} potentials
and respectively the $\Delta N$ as the mass difference $M_\Delta-M$
to generate the strong correlations.
The other potentials are defined in eqs. \eqref{weakpot} and 
\eqref{pncrhodel}-\eqref{ndtranspot}.
In the presence of $\Delta N$ channels the $NN$ potentials must be 
modified to avoid doubly counting the attraction from the channel 
coupling guaranteeing the phase equivalence. It is crucial for the 
result that it is done with high precision. The reduced masses $\mu$ for
the initial $NN$ and $\Delta N$ states are respectively $\mu_{NN}=M/2$
and $\mu_{N\Delta}=MM_\Delta/(M+M_\Delta)$.

For the weak $NN$ interaction we use the DDH weak one-meson exchange
nucleon-nucleon potential ref. \cite{ddh}. In the case of
$pn$-system, the relevant part of the potential is
\begin{align}\label{weakpot}
\hat{V}_{NN}^{\rm PNC}({\bm{r}})=&\frac{ih_\pi^{(1)}g_\pi}{2\sqrt{2}M}
(\bm{\tau}_1\times\bm{\tau}_2)_z(\bm{\sigma}_1+\bm{\sigma}_2)
\cdot\hat{\mathcal{O}}_\pi^-
-\frac{g_\rho}{M}\Bigl[\Bigl(h_\rho^{(0)}\bm{\tau}_1
\cdot\bm{\tau}_2+\frac{h_\rho^{(2)}}{2}[\bm{\tau}_1
\otimes\bm{\tau}_2]_0^{(2)}\Bigr)\times
\nonumber\\
&\Bigl((\bm{\sigma}_1-\bm{\sigma}_2)
\cdot\hat{\mathcal{O}}_\rho^+
+i(1+\chi_\rho)(\bm{\sigma}_1\times\bm{\sigma}_2)
\cdot\hat{\mathcal{O}}_\rho^-\Bigr)
-h_\rho^{(1)}\frac{\hat{\tau}_1^z
-\hat{\tau}_2^z}{2}(\bm{\sigma}_1
+\bm{\sigma}_2)\cdot\hat{\mathcal{O}}_\rho^+\Bigr]
\nonumber\\
&-\frac{g_\omega}{M}\Bigl[h_\omega^{(0)}\Bigl((\bm{\sigma}_1
-\bm{\sigma}_2)\cdot\hat{\mathcal{O}}_\omega^+
+i(1+\chi_\omega)(\bm{\sigma}_1\times\bm{\sigma}_2)
\cdot\hat{\mathcal{O}}_\omega^-\Bigr)
+h_\omega^{(1)}\frac{\hat{\tau}_1^z
-\hat{\tau}_2^z}{2}(\bm{\sigma}_1
+\bm{\sigma}_2)\cdot\hat{\mathcal{O}}_\omega^+\Bigr],
\end{align}
where $\hat{\mathcal{O}}_\alpha^-=[-i\bm{\nabla},Y_{\alpha}(r)]$
and $\hat{\mathcal{O}}_\alpha^+=\{-i\bm{\nabla},Y_{\alpha}(r)\}$,
with $\alpha=\pi,\rho,\omega$ are respectively the commutator and
anticommutator in which the radial functions are
\begin{equation}\label{yuk}
Y_{\alpha}(r)=\frac{e^{-m_\alpha r}}{4\pi r},
\end{equation}
if form factors are not used.
In case monopole form factors of the type
$(\Lambda_\alpha^2-m_\alpha^2)/(\bm{q}^2+m_\alpha^2)$,
which we use here, are included in vertices, the
modified Yukawa functions take the form
\begin{equation}
\label{modyuk} Y_{\alpha}(r)=\frac{e^{-m_\alpha r}}{4\pi r}
-\frac{e^{-\Lambda_\alpha r}}{4\pi}\Bigl(\frac{1}{r}
+\frac{\Lambda_\alpha^2-m_\alpha^2}{2\Lambda_\alpha}\Bigr).
\end{equation}
In eq. \eqref{weakpot} we have neglected the term proportional
to $h_\rho^{(1)'}$ because of the smallness and vagueness of the
coupling and also the irrelevant $\propto(\hat{\tau}_1^z+\hat{\tau}_2^z)$ terms, which do not contribute in $pn$ interaction. 
\begin{table}[tb]
\caption{\label{param}
The weak $\alpha NN$ couplings $h_\alpha^{(i)}$ and $\alpha
N\Delta$ couplings $h_\alpha^{\star(i)}$. The set of the first
three weak couplings are the DDH "best values" \cite{ddh} and
the following five the FCDH "best values" \cite{fcdh}. The weak
couplings are given in units of $10^{-7}$.}
\begin{ruledtabular}
\begin{tabular}{c|ccc|ccccc|ccc}
\multicolumn{5}{c}{DDH~~~}
&\multicolumn{3}{c}{FCDH}\\
&$ h_\alpha^{(0)}$&$
h_\alpha^{(1)}$&$
h_\alpha^{(2)}$&$
h_\alpha^{(0)}$&$
h_\alpha^{(1)}$&$
h_\alpha^{(2)}$&$
h_\alpha^{\star(0)}$&$
h_\alpha^{\star(1)}$&$
g_\alpha$&$\chi_\alpha$&$
\Lambda_\alpha$~(GeV)\\
\hline
$~\pi~$ & - & 4.6 & - & - & 2.7 & - &
- & - &
13.45 & - & 1.2\\
$~\rho~$ & $-11.4$ & $-0.2$ & $-9.5$ &
$-3.8$ & $-0.4$ & $-6.8$ &
7.6 & 7.6 &
2.79 & 3.71 & 1.2\\
$~\omega~$ & $-1.9$ & $-1.1$ & - &
$-4.9$ & $-2.3$ & - &
- & 4.2 &
8.37 & $-0.12$ & 1.2\\
\end{tabular}
\end{ruledtabular}
\end{table}

The PNC transition potential
$(NN\leftrightarrow\Delta N)$ may be derived from the vertex interaction
Hamiltonians \eqref{hampc}-\eqref{pncpidel} of Appendix \ref{appa},
resulting in the potentials \eqref{pncrhodel}-\eqref{pncomegadel} for 
$\rho$-,$\omega$-, and $\pi$-exchanges respectively:
\begin{align}\label{pncrhodel}
V_{\rho N\Delta}^{{\rm PNC}}(\bm{r})
=-\frac{1}{2M}
\Biggl\{&g_\rho \Biggl[
\Bigl(h_\rho^{\star(0)}
+\frac{h_\rho^{\star(1)}}{3}\Bigr)
\bm{T}_1\cdot\bm{\tau}_2
+h_\rho^{\star(1)}\sqrt{\frac{2}{3}}
[\bm{T}_1\otimes\bm{\tau}_2]^{(2)}_0\Biggr]
\bm{S}_1\cdot\hat{\mathcal{O}}_\rho^+
\nonumber\\
&+\Bigg[\Biggl(g_\rho^{\star}h_\rho^{(0)}+
g_\rho\Bigl( h_\rho^{\star(0)}
+\frac{h_\rho^{\star(1)}}{3}\Bigr)\Biggr)
\bm{T}_1\cdot\bm{\tau}_2
+\Bigl(\frac{g_\rho^{\star}h_\rho^{(2)}}{2}
+\sqrt{\frac{2}{3}}g_\rho h_\rho^{\star(1)}
\Bigr)[\bm{T}_1\otimes\bm{\tau}_2]^{(2)}_0
\nonumber\\
&~~~~~+h_\rho^{(1)}g_\rho^\star\hat{T}_{10}\Biggr]
i(1+\chi_\rho)(\bm{S}_1\times\bm{\sigma}_2)
\cdot\hat{\mathcal{O}}_\rho^-
\nonumber\\
&-g_\rho \Biggl[
\Bigl(h_\rho^{\star(0)}
+\frac{h_\rho^{\star(1)}}{3}\Bigr)
\bm{\tau}_1\cdot\bm{T}_2
+h_\rho^{\star(1)}\sqrt{\frac{2}{3}}
[\bm{\tau}_1\otimes\bm{T}_2]^{(2)}_0\Biggr]
\bm{S}_2\cdot\hat{\mathcal{O}}_\rho^+
\nonumber\\
&+\Bigg[\Biggl(g_\rho^{\star}h_\rho^{(0)}+
g_\rho\Bigl( h_\rho^{\star(0)}
+\frac{h_\rho^{\star(1)}}{3}\Bigr)\Biggr)
\bm{\tau}_1\cdot\bm{T}_2
+\Bigl(\frac{g_\rho^{\star}h_\rho^{(2)}}{2}
+\sqrt{\frac{2}{3}}g_\rho h_\rho^{\star(1)}
\Bigr)[\bm{\tau}_1\otimes\bm{T}_2]^{(2)}_0
\nonumber\\
&~~~~~+h_\rho^{(1)}g_\rho^\star\hat{T}_{20}\Biggr]
i(1+\chi_\rho)(\bm{\sigma}_1\times\bm{S}_2)
\cdot\hat{\mathcal{O}}_\rho^-
\Biggr\}+{\rm h.c.},
\end{align}
\begin{align}\label{pncpiondel}
V_{\pi N\Delta}^{{\rm PNC}}(\bm{r})
=&i\frac{h_\pi^{(1)}g_\pi^\star}{2\sqrt{2}M}
\Bigg(
(\bm{T}_1\times\bm{\tau}_2)_0\bm{S}_1
+(\bm{\tau}_1\times\bm{T}_2)_0\bm{S}_2
\Bigg)\cdot\hat{\mathcal{O}}_\pi^-+{\rm h.c.},
\end{align}
and
\begin{align}\label{pncomegadel}
V_{\omega N\Delta}^{{\rm PNC}}(\bm{r})
=&-\frac{g_\omega
h_\omega^{\star(1)}}{2M}
\Bigg[
\Bigl(\hat{T}_{10}\bm{S}_1-\hat{T}_{20}\bm{S}_2\Bigr)
\cdot\hat{\mathcal{O}}_\omega^+\nonumber\\
&+i(1+\chi_\omega)\Bigl(
\hat{T}_{10}(\bm{S}_1
\times\bm{\sigma}_2)
+\hat{T}_{20}(\bm{\sigma}_1\times\bm{S}_2)
\Bigr)
\cdot\hat{\mathcal{O}}_\omega^-\Bigg]
+{\rm h.c.},
\end{align}
where $g_\alpha^\star=\sqrt{72/25}\, g_\alpha$ ($\alpha=\pi,\rho$) are 
the strong meson-$N\Delta$ couplings following from the quark model.
It may be noted that in the case of pion-exchange the $\Delta$ is 
generated only at the strong vertex. 
The $\pi$ and $\rho$ -mediated strong transition potential of the
standard form \cite{brown} is
\begin{align}\label{ndtranspot}
\hat{V}_{N\Delta}^{\pi,\rho}(\bm{r})
=&~\frac{g_\pi g_\pi^\star}{4M^2}
\bm{T}_1\cdot\bm{\tau}_2
(\bm{S}_1\cdot\bm{\nabla})
({\bm{\sigma}}_2\cdot\bm{\nabla})Y_\pi(r)
\nonumber\\
&+\frac{g_\rho g_\rho^\star}{4M^2}(1+\chi_\rho)^2
\bm{T}_1\cdot\bm{\tau}_2
(\bm{S}_1\times\bm{\nabla})
({\bm{\sigma}}_2\times\bm{\nabla})Y_\rho(r)
+(1\leftrightarrow2)+{\rm h.c.},
\end{align}
involving a spin-spin and tensor part.
Note that the strong $N\Delta$ transition potential \eqref{ndtranspot}
is more singular than $r^{-2}$ and thus necessarily requires 
regularization. Therefore, in order to be thoroughly consistent, we
always use the modified Yukawa functions \eqref{modyuk} in the presence
of the $\Delta$.

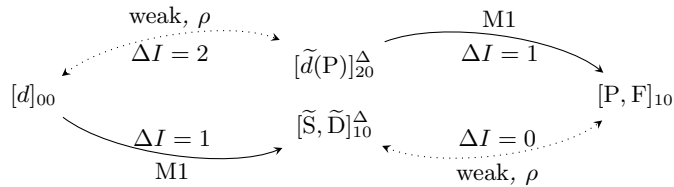
\begin{figure}
    \centering
\begin{tikzpicture}[>=stealth,->,shorten >=2pt,
looseness=.7,auto]
    \matrix [matrix of math nodes,
        column sep={4cm,between origins},
        row sep={0.4cm,between origins}]
        {
            & |(A)| [\widetilde{d}({\rm P})]_{20}^\Delta & \\
            |(B)| [d]_{00}  && |(C)|
            [{\rm P, F}]_{10} \\
            & |(D)| [{\widetilde{\rm S},
            \widetilde{\rm D}}]_{10}^\Delta \\
        };
    \tikzstyle{every node}=[font=\small\itshape]
    \draw (A) to [bend left] (C)
     node [midway,above] {{\rm M1}}
     node [midway,below] {$\Delta I=1$};
    \draw (B) to [<->,dotted,bend left] (A)
     node [midway,above] {{\rm weak}, $\rho$}
     node [midway,below] {$\Delta I=2$};
    \draw (B) to [bend right] (D)
     node [midway,above] {$\Delta I=1$}
     node [midway,below] {{\rm M1}};
    \draw (D) to [<->,dotted,bend right] (C)
     node [midway,above] {$\Delta I=0$}
     node [midway,below] {{\rm weak}, $\rho$};
\end{tikzpicture}
\caption{\label{disintproce}Diagrammatic representation of the 
isospin change $\Delta I$ in the process of the PNC reaction
$\vec{\gamma}d\rightarrow np$ through the direct
$\Delta$-channel. The subscript denotes the isospin state.}
\end{figure}

The whole process of the PNC photodisintegration 
$\vec{\gamma}d\rightarrow np$ is compelled to change the initial (PC bound)
isosinglet state to the final (continuum) isovector state. The
disintegration isospin transition operator allows only the $\Delta I=1$
transitions, with $\langle10\vert\hat{T}_{iz}\vert00\rangle=
\langle20\vert\hat{T}_{iz}\vert10\rangle=
\sqrt{2/3}$ being the only nonzero matrix elements, where $i=1,2$
labels the particle.
Only the isovector mesons can couple to the PC $N\Delta$-vertex, which
automatically excludes the $\omega$-exchange in such amplitudes.
The weak $\pi$-exchange is also excluded, since the total isospin
would not change in processes via $\Delta$ channels. For the same
reason there is no contribution coming from the PNC
%structures related to
$N\Delta\omega$ vertex. In general, all the structures of the PNC
transition potential related to the isospin $\Delta I=1$ change  
are zero, and thus only the $\rho$-exchange occurs in the presence
of the $\Delta$-channel, see fig. \ref{disintproce}.

The potential \eqref{pncrhodel} has only one relevant isospin
transition amplitude $1\leftrightarrow1$ in scattering and basically
two $0\leftrightarrow0$ and $0\leftrightarrow2$ in the deuteron, from
which only the latter one is nonzero.
Following the usual practice in similar calculations, we also choose to
use the DDH couplings in order to be comparable with the corresponding
works.
Unfortunately, there are no published weak $\alpha N\Delta$ 
($\alpha=\pi, \rho, \omega$) couplings corresponding to the weak DDH
$\alpha NN$ couplings.
The "best" values of the weak $\alpha N\Delta$-couplings have been
evaluated in newer analyses \cite{fcdh,despl}. To extract the pure
$\Delta$ effect we shall use
the FCDH values  
\cite{fcdh} for the needed weak $\alpha N\Delta$-couplings,
even though they would not necessarily be entirely consistent with
the other DDH couplings. However, we also study the effect of using
the couplings from the consistent analysis of ref. \cite{fcdh}.

\begin{figure}[h]
\includegraphics[width=4.5cm]{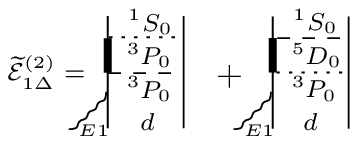}
\caption{\label{correlations}
Second order $\Delta$ corrections at the threshold.}
\end{figure}

It may be noted that,
as a consequence of coupled-channel dynamics, in addition to the
one-meson exchanges depicted in fig. \ref{graphs}
the $\Delta$ channels
may also include possible higher order corrections,
which are naturally taken into account as correlation effects.
For instance, at the threshold, the leading $\Delta$
contributions originate from the once-iterated meson exchange
diagrams presented in fig. \ref{correlations}, where eq. 
\eqref{ndtranspot} is employed to take care of the strong $N\Delta$
transitions.

\subsection{Observables}

\begin{figure}[b]
\begin{minipage}[b]{0.38\linewidth}
\centering
\includegraphics[scale=1.3]{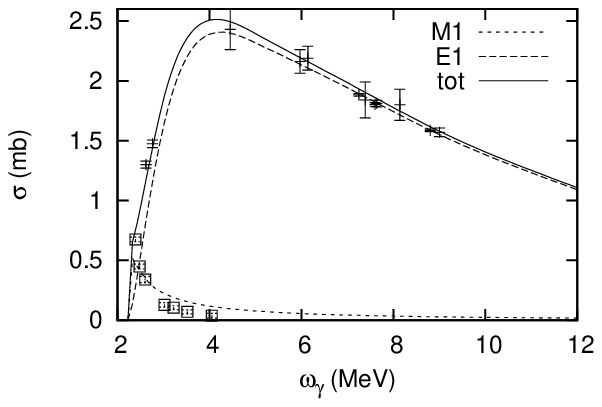}
%\label{fig:figure1}
\end{minipage}
\hspace{2.0cm}
\begin{minipage}[b]{0.48\linewidth}
\centering
\includegraphics[scale=1.3]{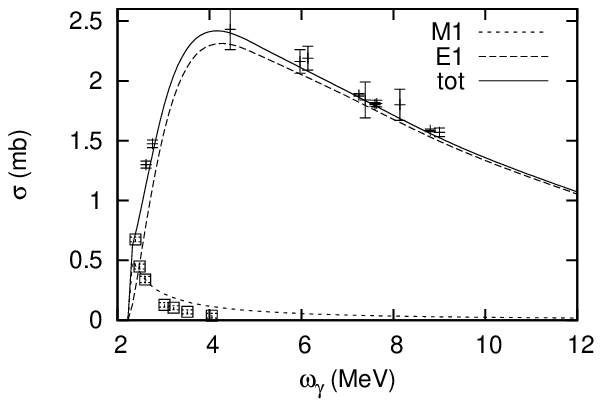}
%\label{fig:figure2}
\end{minipage}
\caption{\label{acs}
The photodisintegration cross-sections of a deuteron
for E1, M1, and their sum given by eq. \eqref{totcs}.
The Reid93 potential is employed in the left-hand side
figure and the A$v_{18}$ potential in
the right-hand side figure. The data points ($\Box$) for M1 are taken
from ref. \cite{m1exp} and ($+$) for the total cross-section from
ref. \cite{csexp2}. The $\omega_\gamma$ is the incident photon laboratory energy.}
\end{figure}

\begin{figure}[tb]
\begin{minipage}[b]{0.38\linewidth}
\centering
\includegraphics[scale=1.3]{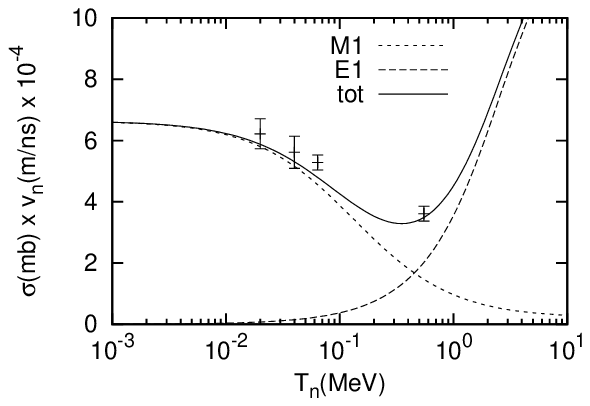}
%\label{fig:figure1}
\end{minipage}
\hspace{2.0cm}
\begin{minipage}[b]{0.48\linewidth}
\centering
\includegraphics[scale=1.3]{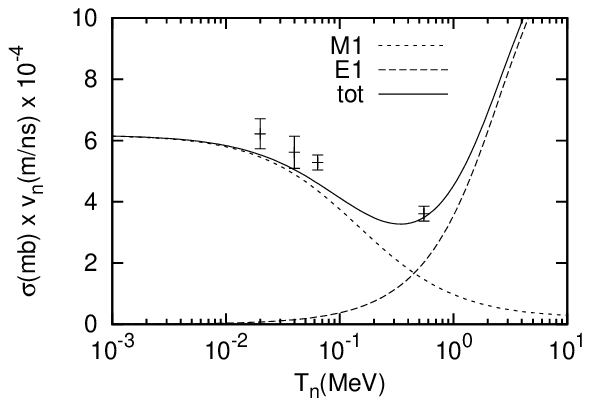}
%\label{fig:figure2}
\end{minipage}
\caption{\label{ncapcs}
The neutron capture cross-sections on a proton
given by eq. \eqref{npcs} (the Reid93 potential is used
in the left and the A$v_{18}$ right).
The data is from refs. \cite{suzuki,nagai}.
The lack of the meson-exchange currents
(mainly the one-pion-exchange current correction)
weakens the
M1 of about 10\% at low energies \cite{riska}.
The neutron speed $v_{\rm n}$ (in the units of
meters/nanosecond) and kinetic energy $T_{\rm n}$
are in the laboratory frame.}
\end{figure}

As a first test we calculate
the total photon absorption cross-section $\sigma_\gamma =
\sigma_++\sigma_-$ for the reaction 
$\gamma d\rightarrow np$, obtained by Fermi's golden rule
summing over the photon helicities and the final spin
states and averaging over the two polarization directions of
the photon and the three possible spin projections of the
deuteron.
Due to the identity of the final state
particles (in the isospin formalism)
the total cross section is also divided by two.
Thus we have
\begin{align}\label{totcs}
\sigma_\gamma=
\frac{2M}{9k}
\sum_{\kappa_d\kappa J}
\Bigl(
|\mathcal{E}^{\kappa J}_{\kappa_d}|^2
+|\mathcal{M}^{\kappa J}_{\kappa_d}|^2
\Bigr),
\end{align}
where $\mathcal{M}^{\kappa J}_{\kappa_d}$ and
$\mathcal{E}^{\kappa J}_{\kappa_d}$ are respectively
the general reduced matrix elements of the magnetic and electric
transitions given in Appendix \ref{appb}. 
The isovector $\mathcal{M}_{{}^3S_1}^{{}^1S_0}$ transition is the 
only non-negligible M1 contribution at low energies and is 
furthermore non-vanishing only in the threshold domain.
The result is in good agreement with experimental data, as
seen in figs. \ref{acs} and \ref{ncapcs}.
In the latter figure we have used the reciprocity
relation
\begin{equation}\label{npcs}
\sigma_{\rm n}=\frac{3}{2}
\Bigl(\frac{\omega_\gamma}{k}\Bigr)^2\sigma_\gamma,
\end{equation}
for the inverse reaction $np\rightarrow\gamma d$. 

%
%             CROSS-SECTIONS PHOTODISINTEGRATION
%
%
%
%             CROSS-SECTIONS NEUTRON CAPTURE
%

The relevant asymmetry observable for the reaction
$\vec{\gamma}d\rightarrow np$ may be expressed in terms of the
deuteron photodisintegration helicity cross sections $\sigma_\lambda$
as
\begin{equation}\label{asym}
\mathcal{A}_{\vec{\gamma}}
=\frac{\sigma_+-\sigma_-}{\sigma_++\sigma_-}
=\frac{2{\rm Re}
\sum_{\kappa\kappa'J}\sum_{\kappa_d\kappa_d'}
\Bigl(
\mathcal{M}^{\kappa J\ast}_{\kappa_d}
\widetilde{\mathcal{E}}^{\kappa'J}_{\kappa_d'}
+\mathcal{E}^{\kappa J\ast}_{\kappa_d}
\widetilde{\mathcal{M}}^{\kappa'J}_{\kappa_d'}\Bigr)}
{\sum_{\kappa\kappa_dJ}\Bigl(
|\mathcal{M}^{\kappa J}_{\kappa_d}|^2
+|\mathcal{E}^{\kappa J}_{\kappa_d}|^2
+|\widetilde{\mathcal{E}}^{\kappa J}_{\kappa_d}|^2
+|\widetilde{\mathcal{M}}^{\kappa J}_{\kappa_d}|^2\Bigr)}\, .
\end{equation}
Here and later on, the parity admixed (weak)
wavefunctions and amplitudes generated by the weak
nuclear force will be tilded for clarity.
Naturally, the absolute squares of the weak
amplitudes in the denominator of eq. \eqref{asym} may as well
be ignored due to their diminutive size. 
Specializing to low partial waves they reduce to the results given
{\it e.g.} in refs. \cite{liu} and \cite{haid}.
The asymmetry may also be written following the notation fixed in
fig. \ref{graphs}  as
\begin{equation}\label{asymmetry}
\mathcal{A}_k=
\frac{2{\rm Re}\Bigl[\mathcal{M}_1^\ast
\widetilde{\mathcal{E}}_1+\sum_{i=2}^k
\mathcal{E}_i^\ast\widetilde{\mathcal{M}}_i\Bigr]}
{|\mathcal{M}_1|^2+\sum_{i=2}^k|\mathcal{E}_i|^2},
\end{equation}
where the index $i$ denotes the final continuum channel.
The total asymmetry $\mathcal{A}_{\vec{\gamma}}$ with $S$ and $P$
waves is achieved when $k=4$.

The magnetic dipole effect is dominant at the threshold where
the asymmetry  \eqref{asymmetry} reduces to a simple form
$\mathcal{A}_{\vec{\gamma}}^{\rm thr.}(k\rightarrow0)\leadsto
2{\rm Re}[\widetilde{\mathcal{E}}_1/
\mathcal{M}_1]$,
which is explicitly given by eqs. \eqref{e1} and \eqref{m1}
as
\begin{equation}\label{thrasym}
\mathcal{A}_{\vec{\gamma}}^{\rm thr.}\approx 2{\rm Re}
\Biggl[\frac{\frac{i}{\sqrt{3}}
\int drr\widetilde{\mathcal{U}}^{(+)}_{{}^3P_0}(k,r)
\Bigl(\mathcal{D}_{{}^3S_1}(r)-\sqrt{2}
\mathcal{D}_{{}^3D_1}(r)\Bigr)
-i\int drr\mathcal{U}^{(+)}_{{}^1S_0}(k,r)
\widetilde{\mathcal{D}}_{{}^1P_1}(r)}
{-\frac{\mu_v\sqrt{3}}{M}
\int dr\mathcal{U}^{(+)}_{{}^1S_0}(k,r)
\mathcal{D}_{{}^3S_1}(r)}\Biggr].
\end{equation}
It is important to
note that eq. \eqref{thrasym} arises from the spin changing
PNC interaction, see fig. \ref{graphs}, and thus does not include
the PNC pion exchange. Therefore, 
contributions from heavier vector meson exchanges (and possibly
$\Delta$) are maximized.
The low-energy limit \eqref{thrasym} coincides with the photon 
polarization in the time-reversed reaction $np\rightarrow\vec{\gamma}d$
for thermal neutrons. 

\begin{figure}[tb]
\includegraphics[width=11.cm]{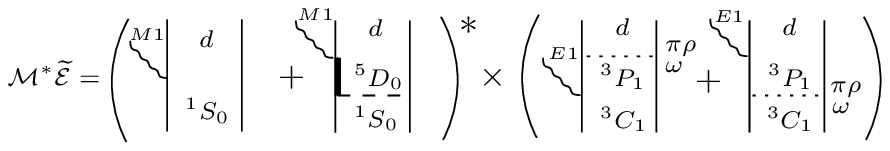}
\caption{\label{graph2}
Graphical representation for the PNC reaction
$\vec{n}p\rightarrow\gamma d$ with
thermal neutrons.}
\end{figure}

The asymmetry from the radiative capture $\vec{n}p\rightarrow\gamma d$
of longitudinally polarized thermal neutrons in hydrogen is also
calculated.  The appropriate scattering wavefunctions depending on 
the spin magnetic quantum numbers ($m_n$ and $m_p$) of the neutron and
proton are obtained by expanding eq. \eqref{scatwf}.
The wavefunctions (assuming the $z$-axis to be along the direction
of $\bm{k}$) become
\begin{align}\label{scatwf2}
\psi_{m_nm_p}^{(-)}(k,\bm{r})=&
\frac{\sqrt{4\pi}}{kr}
\sum_{\kappa'\kappa}\sum_{JM_S}
i^{L}\sqrt{2L+1}
\langle{\textstyle\frac{1}{2}}m_n
{\textstyle\frac{1}{2}}m_p\vert SM_S\rangle
\times\nonumber\\
&\langle L0SM_S\vert JM_S\rangle
\mathcal{U}_{\kappa\kappa'}^{J(-)}(k,r)
\mathcal{Y}^{L'S'}_{JM_S}(\hat{\bm{r}})
(-1)^{T'+1}\vert T'0\rangle.
\end{align}
Since this is a time-reversed process to the deuteron disintegration,
the amplitudes are of the form
$\overline{{}_{\mathcal{D}}\langle M_d|
\hat{H}^{\lambda\dagger}_{{\rm e.m.}}
|k;m_nm_p\rangle^{(-)}}$.
The asymmetry observable, given in terms of spin differential
neutron capture cross sections $d\sigma_{m_n}/d\Omega$ with
the neutron polarization $m_n$ and further in terms of the reduced
matrix elements approximated for the thermal neutrons, reads
$\mathcal{A}_{\vec{n}}(\theta)=\mathcal{A}_{\vec{n}}\cos\theta$ with
\begin{align}\label{asymneu}
&\mathcal{A}_{\vec{n}}=
\frac{d\sigma_{+\frac{1}{2}}
-d\sigma_{-\frac{1}{2}}}{d\sigma_{+\frac{1}{2}}
+d\sigma_{-\frac{1}{2}}}\approx
\sqrt{2}{\rm Re}
\Bigl[
\frac{\widetilde{\mathcal{E}}}{\mathcal{M}}
\Bigr]=
\nonumber\\
&\sqrt{2}{\rm Re}
\Biggl[\frac{
i\int drr\widetilde{\mathcal{D}}_{{}^3P_1}(r)
\Bigl(\mathcal{U}^{(+)}_{{}^3S_1}(k,r)
+\frac{1}{\sqrt{2}}\mathcal{U}^{(+)}_{{}^3D_1}(k,r)
\Bigr)
-i\int drr\Bigl(
\mathcal{D}_{{}^3S_1}(r)
+\frac{1}{\sqrt{2}}\mathcal{D}_{{}^3D_1}(r)
\Bigr)\widetilde{\mathcal{U}}_{{}^3P_1}^{(+)}(k,r)}{
\frac{\sqrt{3}\mu_v}{M}
\int dr\mathcal{D}_{{}^3S_1}(r)
\mathcal{U}_{{}^1S_0}^{(+)}(k,r)}\Biggr],
\end{align}
where $\theta$ is the angle between the momenta of the incident 
neutron and emitted photon. 
Figure \ref{graph2} presents this threshold result diagrammatically.
Contrary to eq. \eqref{thrasym}, the
$\mathcal{A}_{\vec{n}}$ is dominated by the spin conserving weak
interaction and therefore the PNC one pion exchange prevails.
Contrary to $\vec{\gamma}d\rightarrow np$ in this reaction the $\Delta$-corrections are due to 
$\pi$, $\rho$, and $\omega$ exchanges.

\section{\label{sec3}Results and Discussion}

\begin{figure}[tb]
\includegraphics[width=8.5cm]{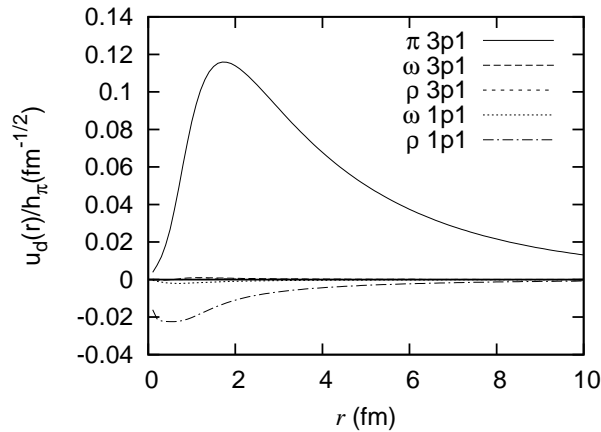}
\caption{\label{difmesddhunmod}
Contributions of different mesons in different
partial waves without the $\Delta$ using the DDH
couplings, the A$v_{18}$ potential, and 
unmodified Yukawa functions.}
\end{figure}

Utilizing the different model complexes outlined in the previous
section, we now proceed to calculate the
asymmetries $\mathcal{A}_{\vec{\gamma}}$ for the reaction
$\vec{\gamma}d\rightarrow np$ as a
function of photon laboratory energies varying from the deuteron
breakup threshold to 10 MeV and $\mathcal{A}_{\vec{n}}$
in the radiative capture of thermal neutrons (25 meV) 
$\vec{n}p\rightarrow\gamma d$.
It is also worth noticing that the photon polarization
$\mathcal{P}_{\vec{\gamma}}$ of the reaction 
$np\rightarrow\vec{\gamma}d$ with thermal
neutrons is in principle the same as the observable of the
time-reversed reaction at threshold 
$\mathcal{P}_{\vec{\gamma}} 
\approx \mathcal{A}_{\vec{\gamma}}(\omega_\gamma^{\rm th.})$.
We also compare in some detail the asymmetries with and
without the effects of the virtual $\Delta$-isobar.  
\begin{figure}[tb]
\begin{minipage}[b]{0.38\linewidth}
\centering
\includegraphics[scale=1.3]{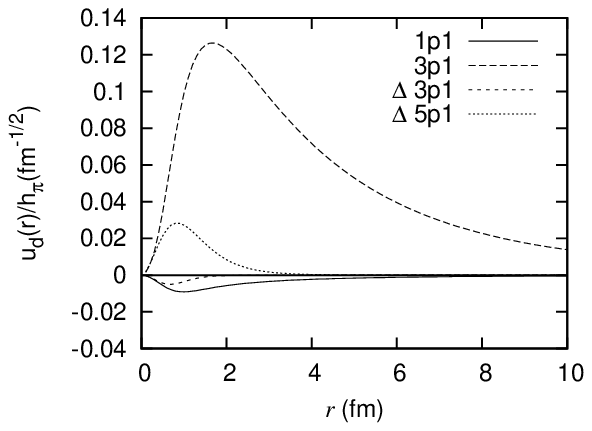}
%\label{dwfddh93}
\end{minipage}
\hspace{2.0cm}
\begin{minipage}[b]{0.48\linewidth}
\centering
\includegraphics[scale=1.3]{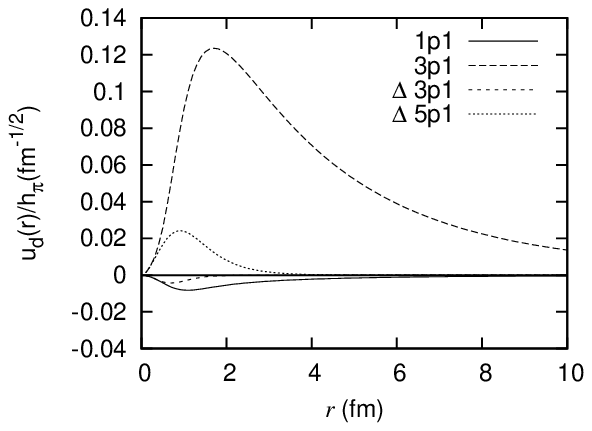}
%\label{dwffcdh18}
\end{minipage}
\caption{\label{extradwf}
The PNC deuteron components (divided by the DDH $h^{(1)}_\pi$)
including the $\Delta$ with the DDH couplings and modified 
Yukawa functions. The Reid93 potential is used in the left 
panel and the A$v_{18}$ in the right.}
\end{figure}
\begin{figure}[tb]
\begin{minipage}[b]{0.38\linewidth}
\centering
\includegraphics[scale=1.3]{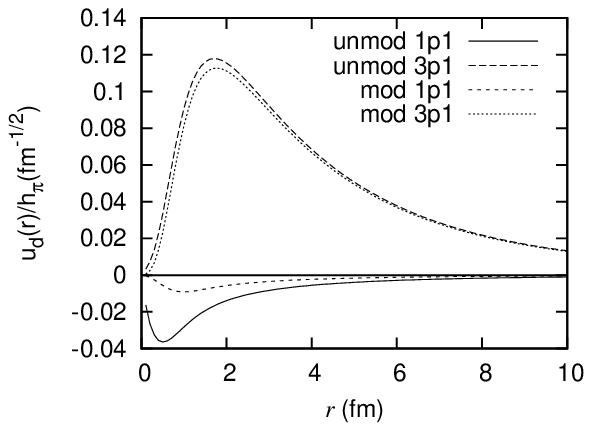}
%\label{dwfddh93}
\end{minipage}
\hspace{2.0cm}
\begin{minipage}[b]{0.48\linewidth}
\centering
\includegraphics[scale=1.3]{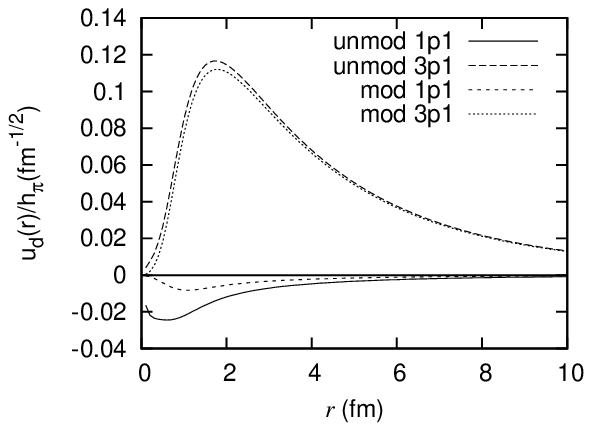}
%\label{dwffcdh18}
\end{minipage}
\caption{\label{dwfddhmodunmod}
The same as fig. \ref{extradwf}, but without the $\Delta$.
The components are compared using both unmodified and modified 
Yukawa functions. The result (unmodified) gained with A$v_{18}$
potential is the same as in ref. \cite{haid}.}
\end{figure}
In figs. \ref{extradwf} and \ref{dwfddhmodunmod} we show the PNC
deuteron components relevant for the reactions discussed in this
paper with and without form factors and also including $\Delta N$
admixture. Two phenomenological potentials are used, the updated 
Reid soft core \cite{reid93} and Argonne $v_{18}$ \cite{av18}
potentials. By far, the largest PNC component is ${}^3P_1$, which 
mainly arises from pion exchange as seen in fig. \ref{difmesddhunmod}. 
However, as was seen in sect. \ref{sec2} and 
fig. \ref{graphs}, this does not contribute to PNC photoabsorption 
into the lowest ${}^1S_0$ partial wave and, consequently, is of 
minor importance at threshold. The ${}^1P_1$ component arises from 
shorter ranged vector meson exchanges, notably $\rho$ and is smaller 
by an order of magnitude. The pion generated part ${}^3P_1$ is 
relatively model independent (except for the weak pion coupling) 
whereas in ${}^1P_1$ some short-range dependence can be seen, if the 
PNC potentials are not moderated by a form factor. Also it can be noted 
$\Delta N$ components can be of the same order as ${}^1P_1$ although 
of shorter range. Therefore, it is plausible that their effect could 
be, in principle, appreciable and should be considered. 
Further, from fig. \ref{difmesddhunmod} it is obvious that with the 
standard DDH couplings $\omega$ exchange would be negligible and 
also $\rho$ in the pion dominated ${}^3P_1$ component. Figure 
\ref{extrafcdh} shows another choice of weak couplings and the 
dependence on these is significant. The FCDH parametrization 
gives smaller PNC in comparison with fig. \ref{extradwf}.
\begin{figure}[tb]
\includegraphics[width=8.5cm]{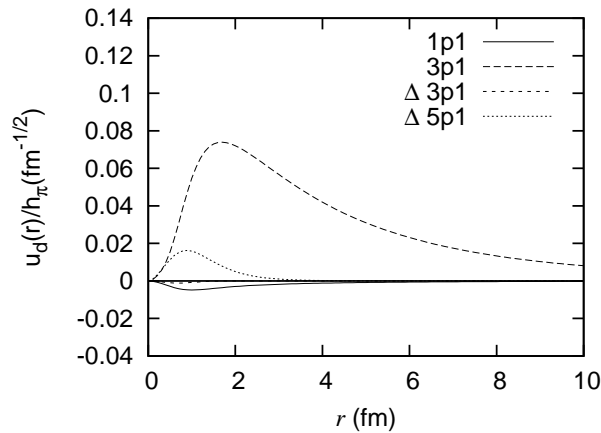}
\caption{\label{extrafcdh}
The same as fig. \ref{extradwf}. The components (divided by the 
DDH $h^{(1)}_\pi$) with the A$v_{18}$ potential and the FCDH couplings}.
\end{figure}
Even though we have compromised to use the FCDH values, the other 
available weak coupling set with $\Delta$ by Desplanques 
\cite{despl} is alike and should give a similar result. 
Otherwise, if the weak couplings are the same, our PNC $NN$ 
wavefunctions are not in contradiction with those displayed in refs. \cite{haid,schia,fuji}.

\begin{figure}[tb]
\begin{minipage}[b]{0.38\linewidth}
\centering
\includegraphics[scale=1.3]{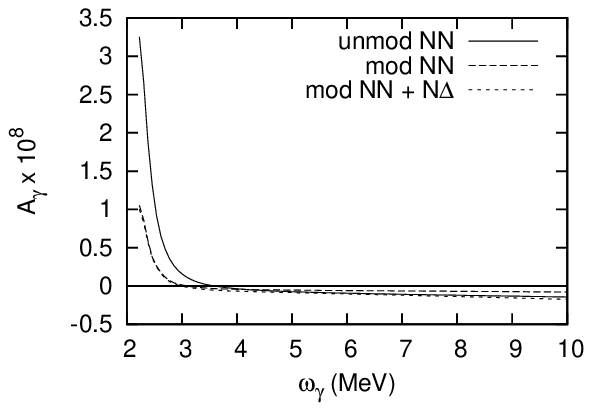}
%\label{fig:figure1}
\end{minipage}
\hspace{2.0cm}
\begin{minipage}[b]{0.48\linewidth}
\centering
\includegraphics[scale=1.3]{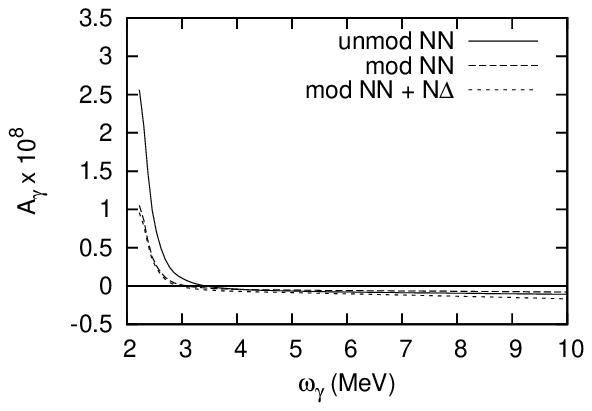}
%\label{fig:figure2}
\end{minipage}
\caption{\label{asymddhmodunmod}
The total asymmetry given by eq. \eqref{asymmetry} and the DDH 
couplings with unmodified and modified Yukawa functions without
(NN) and with the $\Delta$ excitation (NN+N$\Delta$). The Reid93
left and the A$v_{18}$ right.}
\end{figure}

We now consider the asymmetry $\mathcal{A}_{\vec{\gamma}}$, covering 
the photon energies from threshold to 10 MeV. Figure \ref{asymddhmodunmod} illustrates separately the asymmetry of the
"basic model" (by which we mean the commonly used model in the PNC
calculations, which employs the DDH and strong interaction parameter
values of table \ref{param} without form factors) and the effect of the
form factors and the $\Delta$ employing Reid93 and A$v_{18}$ potentials. In
the case of the "basic model", the asymmetry is nearly 30\% bigger with
Reid93 at threshold. This difference is explained by the distinctly
bigger PNC $^1P_1$ component in the deuteron produced by Reid93
potential, as seen in fig. \ref{dwfddhmodunmod}. However, with form 
factors in the PNC potentials the short-range differences are minimized
and the results become graphically indistinguishable. 

These points are further featured in tables \ref{tab3} (last three columns)
and \ref{tab4} in the threshold limit 
$\omega_\gamma^{{\rm lab}}\rightarrow 2.22592$ MeV. 
%
%            TABLES THRESHOLD ASYMMETRY VALUES
%
%
\begin{table}[tb]
\caption{\label{tab3}
Contributions of different mesons to the asymmetry
$\mathcal{A}_{\vec{\gamma}}=
\sum_\alpha\mathcal{A}_{\vec{\gamma}}^\alpha$, 
in units of $10^{-8}$, at threshold
without $\Delta$-excitation
using the unmodified Yukawa functions eq.
\eqref{yuk}. 
The threshold  
asymmetries are split into two pieces
$\mathcal{A}_{\vec{\gamma}}^\alpha=
\mathcal{A}^{{}^3\widetilde{P}_0\alpha}_d+
\mathcal{A}^{{}^1S_0\alpha}_{{}^1\widetilde{P}_1}$,
see eq. \eqref{thrasym}, where the indices characterize 
the E1 transition: subscripts the initial and superscripts 
the final states ($d$ for the PC part).
The $\mathcal{A}_{\vec{\gamma}}^\pi=0$ in every case.}
\begin{tabular}{|c||ccc|ccc|ccc|}
\hline\hline
Model
&$\mathcal{A}^{{}^3\widetilde{P}_0\rho}_d$
&$\mathcal{A}^{{}^1S_0\rho}_{{}^1\widetilde{P}_1}$
&$\mathcal{A}_{\vec{\gamma}}^\rho$
&$\mathcal{A}^{{}^3\widetilde{P}_0\omega}_d$
&$\mathcal{A}^{{}^1S_0\omega}_{{}^1\widetilde{P}_1}$
&$\mathcal{A}_{\vec{\gamma}}^\omega$
&$\mathcal{A}^{{}^3\widetilde{P}_0}_d$
&$\mathcal{A}^{{}^1S_0}_{{}^1\widetilde{P}_1}$
&$\mathcal{A}_{\vec{\gamma}}$\\
\hline
{\rm A$v_{18}$ \& DDH}
&~-0.85~&~3.51~&~2.66~
&~-0.39~&~0.29~&~-0.10~
&~-1.24~&~3.80~&~2.56~\\
{\rm Reid93 \& DDH}
&~-0.84~&~4.10~&~3.26~
&~-0.38~&~0.38~&~0.00~
&~-1.22~&~4.48~&~3.26~\\
{\rm A$v_{18}$ \& FCDH}
&~0.41~&~1.17~&~1.58~
&~-1.00~&~0.76~&~-0.24~
&~-0.59~&~1.93~&~1.34~\\
{\rm Reid93 \& FCDH }
&~0.40~&~1.36~&~1.76~
&~-0.99~&~0.99~&~0.00~
&~-0.59~&~2.35~&~1.76~\\
\hline\hline
\end{tabular}
\end{table}
\begin{table}[tb]
\caption{\label{tab4}
Asymmetries
$\mathcal{A}_{\vec{\gamma}}$
 and
$\mathcal{A}_{\vec{\gamma}}^\Delta$, 
in units of $10^{-8}$, respectively without and with 
$\Delta$-excitation using the modified Yukawa functions 
eq. \eqref{modyuk}.} \begin{tabular}{|c||ccc|ccc|}
\hline\hline
Model
&$\mathcal{A}^{{}^3\widetilde{P}_0}_d$
&$\mathcal{A}^{{}^1S_0}_{{}^1\widetilde{P}_1}$
&$\mathcal{A}_{\vec{\gamma}}$
&$\mathcal{A}^{{}^3\widetilde{P}_0\Delta}_d$
&$\mathcal{A}^{{}^1S_0\Delta}_{{}^1\widetilde{P}_1}$
&$\mathcal{A}_{\vec{\gamma}}^\Delta$\\
\hline
{\rm A$v_{18}$ \& DDH }
&~-0.69~&~1.74~&~1.05~&~-0.73~&~1.69~&~0.96~\\
{\rm Reid93 \& DDH }
&~-0.69~&~1.83~&~1.14~&~-0.73~&~1.77~&~1.04~\\
{\rm A$v_{18}$ \& FCDH }
&~-0.33~&~0.95~&~0.62~&~-0.27~&~0.92~&~0.65~\\
{\rm Reid93 \& FCDH }
&~-0.33~&~1.03~&~0.70~&~-0.28~&~1.00~&~0.72~\\
\hline\hline
\end{tabular}
\end{table}
Short-range correlation 
differences seem appreciable only in the transition from the PNC deuteron
component ${}^1\widetilde{P}_1$ to the ${}^1S_0$ final state without form 
factors (table \ref{tab3}), while with form factors the difference is hardly
10 \%. The effect of the form factors themselves in PNC heavy meson exchanges is, however, a dramatic decrease to less than half of the original value of
the asymmetry $\mathcal{A}_{\vec{\gamma}}$. 
\begin{figure}[tb]
\includegraphics[width=8.5cm]{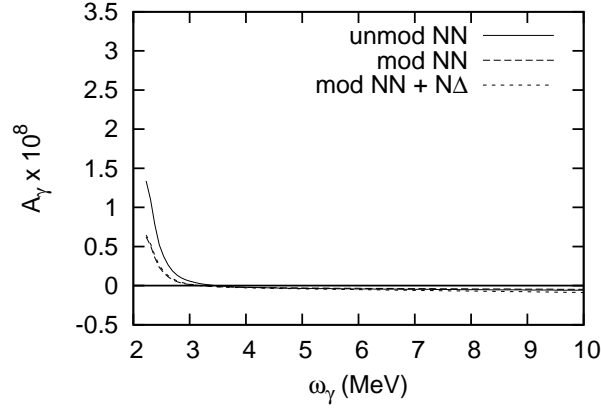}
\caption{\label{asymfcdhmodunmod}
The total asymmetry given by eq. \eqref{asymmetry},
the A$v_{18}$ potential, and the FCDH couplings 
with unmodified and modified Yukawa functions 
without (NN) and with the $\Delta$ excitation (NN+N$\Delta$).}
\end{figure}
\begin{figure}[tb]
\begin{minipage}[b]{0.38\linewidth}
\centering
\includegraphics[scale=1.3]{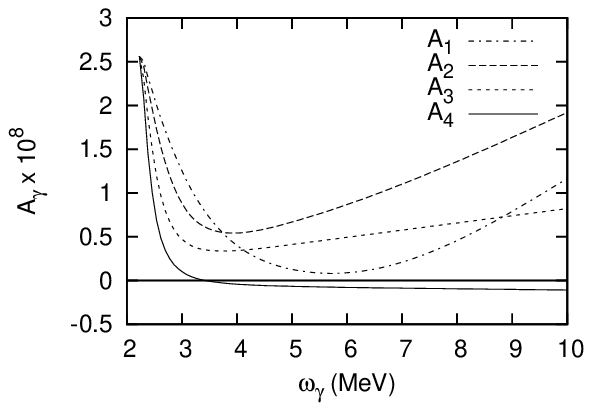}
%\label{fig:figure1}
\end{minipage}
\hspace{2.0cm}
\begin{minipage}[b]{0.48\linewidth}
\centering
\includegraphics[scale=1.3]{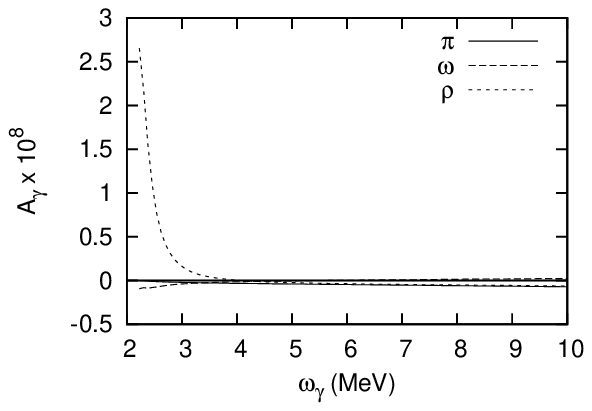}
%\label{fig:figure2}
\end{minipage}
\caption{\label{cumu}
Cumulative behaviour of the asymmetry eq. \eqref{asymmetry}
in the left-hand side panel. Single meson $\pi, \omega,$
and $\rho$ contributions to the asymmetry in the right-hand
side panel. Both figures are plotted employing the DDH
couplings, unmodified Yukawa functions and A$v_{18}$ potential 
without $\Delta$ excitation.}
\end{figure}
In general the FCDH results are much smaller that DDH with and without
form factors as seen in fig. \ref{asymfcdhmodunmod} and Tables.
The threshold result $\mathcal{A}_{\vec{\gamma}}=2.56\times10^{-8}$ 
gained with the "basic model" and the A$v_{18}$ potential argrees with 
most of the existing calculations and perfectly with those of using the
same model \cite{haid,liu}. 
Even though the $N\Delta$ components in the 
deuteron appeared significant in fig. \ref{extradwf}, their effect
in $\mathcal{A}_{\vec{\gamma}}$ remains negligible at low energies being
about 10\% decrease for DDH at threshold and practically null for FCDH. 

Figure \ref{cumu} shows the importance of the $P$-wave continuum states 
above threshold as a cumulative behaviour of eq. \eqref{asymmetry}.
In the same figure are also shown the contributions of different mesons.
As the photon energy increases, the asymmetry decreases steeply within 
the energies up to of about 1 MeV above the threshold. 
This energy region is dominated by heavy meson exchanges, mainly the 
$\rho$. Increasing the energy further, the asymmetry quickly settles 
down near to zero continuing its gentle monotonic decline through zero
somewhere between 3 and 4 MeV, after which the pion starts to take the
dominance.

At low energies the pion contribution to $\mathcal{A}_{\vec{\gamma}}$ is
small and, in our model and expressions, vanishes in the threshold limit
(there is a negligible contribution higher order in 
$\bm{k}_\gamma$ arising from the spin current \cite{schia,liu}). 
This is not so for radiative capture of longitudinally 
polarized neutrons, where it can participate in spin conserving PNC 
transitions ${}^3S_1+{}^3D_1\leftrightarrow{}^3P_1$ as depicted in fig.
\ref{graph2} and expressed in eq. \eqref{asymneu}. As seen in table 
\ref{tab6}, contrary to $\mathcal{A}_{\vec{\gamma}}$, now the vector
meson contributions are negligible. (It may still be of interest to note
that now due to the spin conserving couplings $h_\alpha^{(1)}$ $\omega$
is more important than $\rho$.) PNC in both bound states and continuum
is about equal in magnitude and their contributions add constructively.
Also in this case the value of $\mathcal{A}_{\vec{n}}=-5.39\times10^{-8}$
by the "basic model" and the A$v_{18}$ potential is in agreement with
results of most authors and especially ref. \cite{haid} in which the 
wavefunctions of the same model are used.
\begin{table}[tb]
\caption{\label{tab6}
Contributions of different mesons to the
$\mathcal{A}_{\vec{n}}=\sum_\alpha\mathcal{A}_{\vec{n}}^\alpha$, 
in units of $10^{-8}$, without $\Delta$-excitation
using the unmodified Yukawa functions eq. \eqref{yuk}.
Furthermore, as in table \ref{tab3}, the $\mathcal{A}_{\vec{n}}^\alpha$ 
is a sum of PNC scattering 
$\mathcal{A}_{{}^3\widetilde{P}_1}^{d\alpha}$ and bound 
$\mathcal{A}_{{}^3C_1}^{{}^3\widetilde{P}_1\alpha}$
contributions corresponding to eq. \eqref{asymneu}.}
\begin{tabular}{|c||ccc|ccc|ccc|ccc|}
\hline\hline
Model
&$\mathcal{A}_{{}^3\widetilde{P}_1}^{d\pi}$
&$\mathcal{A}_{{}^3C_1}^{{}^3\widetilde{P}_1\pi}$
&$\mathcal{A}_{\vec{n}}^\pi$
&$\mathcal{A}_{{}^3\widetilde{P}_1}^{d\rho}$
&$\mathcal{A}_{{}^3C_1}^{{}^3\widetilde{P}_1\rho}$
&$\mathcal{A}_{\vec{n}}^\rho$
&$\mathcal{A}_{{}^3\widetilde{P}_1}^{d\omega}$
&$\mathcal{A}_{{}^3C_1}^{{}^3\widetilde{P}_1\omega}$
&$\mathcal{A}_{\vec{n}}^\omega$
&$\mathcal{A}_{{}^3\widetilde{P}_1}^d$
&$\mathcal{A}_{{}^3C_1}^{{}^3\widetilde{P}_1}$
&$\mathcal{A}_{\vec{n}}$\\
\hline
{\rm A$v_{18}$ \& DDH}
&~-2.85~&~-2.51~&~-5.36~
&~0.00~&~0.00~&~0.00~
&~-0.01~&-0.02~&~-0.03~
&~-2.86~&~-2.53~&~-5.39~\\
{\rm Reid93 \& DDH}
&~-2.85~&~-2.37~&~-5.22~
&~0.02~&~0.00~&~0.02~
&~-0.03~&~-0.01~&~-0.04~
&~-2.86~&~-2.40~&~-5.26~\\
{\rm A$v_{18}$ \& FCDH}
&~-1.67~&~-1.47~&~-3.14~
&~0.00~&~0.00~&~0.00~
&~-0.02~&~-0.05~&~-0.07~
&~-1.69~&~-1.52~&~-3.21~\\
{\rm Reid93 \& FCDH }
&~-1.68~&~-1.39~&~-3.07~
&~0.01~&~0.00~&~0.01~
&~-0.02~&~-0.05~&~-0.07~
&~-1.69~&~-1.44~&~-3.13~\\
\hline\hline
\end{tabular}
\end{table}
Due to the long range of the pion and low energies the form factor effect
is only a 2-3 \% decrease in $\mathcal{A}_{\vec{n}}$.   
The $\Delta$ excitation contributes another 5 \% in both continuum and
bound states, but these effects cancel off.

In summary, we have calculated the asymmetries $\mathcal{A}_{\vec{\gamma}}$
and $\mathcal{A}_{\vec{n}}$ in polarized photon absorption and in radiative
capture of polarized neutrons close to threshold. The results are mutually
complementary in the sense that $\mathcal{A}_{\vec{\gamma}}$ is dominated
by vector meson exchange, while $\mathcal{A}_{\vec{n}}$ is pion dominated.
The dominances are more than an order of magnitude with the minor effect
being negligible. Also the $\Delta$ effects are small. In addition, very
soon above threshold $\mathcal{A}_{\vec{\gamma}}$ becomes probably too
small to be experimentally informative. In turn $\mathcal{A}_{\vec{n}}$
can give some limits to the weak pion coupling provided the error can be
pushed below $10^{-8}$ as hoped for the NPDGamma experiment \cite{npdgamma}.
As for $\mathcal{A}_{\vec{\gamma}}$, being sensitive to short range effects,
it could carry information on low energy constants of chiral perturbation
theory. 
\begin{acknowledgements}
T. M. P. would like to thank Vilho, Yrj\"{o} and Kalle
V\"{a}is\"{a}l\"{a} Foundation for the financial support 
of this work. This work was also supported by the Academy 
of Finland and DAAD researcher exchange grants 121892 and 
139512. We thank Institut f\"{u}r Kernphysik of 
Forschungszentrum J\"{u}lich for kind hospitality. 
\end{acknowledgements}
\appendix
\section{}\label{appa}
The parity conserving and nonconserving effective Hamiltonians
(for PNC see \cite{fcdh}) for
the $NN$ and $\Delta N$ vertices are given by
\begin{equation}\label{hampc}
\mathcal{H}_{NN\rho}^{{\rm PC}}=g_\rho\bar{\psi}
\Bigl(\gamma^\mu+\frac{i\chi_\rho}{2M}\sigma^{\mu\nu}q_\nu
\Bigr)\bm{\tau}\psi\cdot\bm{\rho}_\mu,
\end{equation}
\begin{equation}\label{hampnc}
\mathcal{H}_{NN\rho}^{{\rm PNC}}=\bar{\psi}
\Bigl[h_\rho^{(0)}\bm{\tau}\cdot\bm{\rho}_\mu
+h_\rho^{(1)}\rho_\mu^0
+\frac{h_\rho^{(2)}}{2\sqrt{6}}
(3\hat{\tau}_0\rho_\mu^0-\bm{\tau}\cdot\bm{\rho}_\mu)
\Bigr]\gamma^\mu\gamma_5\psi,
\end{equation}
\begin{equation}\label{hampncdel}
\mathcal{H}_{\Delta N\rho}^{{\rm PNC}}
=h_\rho^{\star(0)}\bar{\psi}
\Psi^{\mu i}\rho_{\mu i}+
h_\rho^{\star(1)}\bar{\psi}
\Psi^{\mu 0}\rho_{\mu 0}+{\rm h.c.},
\end{equation}
\begin{equation}\label{hampcdel}
\mathcal{H}_{\Delta N\rho}^{{\rm PC}}
=i\frac{f_\rho^\star}{m_\rho}\bar{\psi}
\gamma_5\gamma^\nu
\bm{T}\Psi^{\mu}\cdot
(\partial_\nu\bm{\rho}_\mu
-\partial_\mu\bm{\rho}_\nu)
+{\rm h.c.},
\end{equation}
\begin{equation}\label{pcome}
\mathcal{H}_{NN\omega}^{{\rm PC}}=g_\omega\bar{\psi}
\Bigl(\gamma^\mu+\frac{i\chi_\omega}{2M}\sigma^{\mu\nu}q_\nu
\Bigr)\psi\omega_\mu,
\end{equation}
\begin{equation}\label{pncome}
\mathcal{H}_{N\Delta\omega}^{{\rm PNC}}
=h_\omega^{\star(1)}\bar{\psi}
\Psi^{\mu 0}\omega_{\mu}+{\rm h.c.},
\end{equation}
\begin{equation}\label{pncpi}
\mathcal{H}_{NN\pi}^{{\rm PNC}}=
\frac{h_\pi^{(1)}}{\sqrt{2}}\epsilon_{ij0}
\bar{\psi}\hat{\tau}_i\hat{\pi}_j\psi,
\end{equation}
\begin{equation}\label{pncpidel}
\mathcal{H}_{N\Delta\pi}^{{\rm PC}}=
\frac{f_\pi^\star}{m_\pi}\bar{\psi}
\Psi^{\mu i}\partial_\mu\hat{\pi}_i+{\rm h.c.}.
\end{equation}
where $f_\rho^\star/m_\rho
=g_\rho^\star(1+\chi_\rho)/2M$ and $\Psi^{\mu}$ is the
Rarita-Schwinger vector-spinor field.
\section{}\label{appb}
In this appendix, we give explicitly the reduced matrix elements
of the Hamiltonian eq. \eqref{emham} expressed in terms of Wigner
coefficients. The factor $N=\sqrt{\alpha\pi\omega_\gamma/2}$
in the reduced matrix elements \eqref{e1}-\eqref{md1} simplifies 
away in the asymmetry observables. The time-reversed
adjoint electromagnetic amplitudes  differ from
the originals only by the phase factor of $-(-1)^{J}$,
which cancels out in the absolute squares and interference.
The reduced $NN$ matrix elements for the electric E1 transition are
(with shorthand notations
$\hat{J}=2J+1$, $\hat{S}=2S+1$, and $\hat{L}=2L+1$)
\begin{align}\label{e1}
\mathcal{E}^{\kappa'J}_{\kappa_d}(k)
&=iN(-1)^{S'}
\sqrt{
3\hat{J}\hat{L}'\hat{L}_d}
\left\{
\begin{array}{ccc}
L' & J & S' \\
1 & L_d & 1 \\
\end{array}\right\}
\left(
\begin{array}{ccc}
L' & 1 & L_d \\
0 & 0 & 0 \\
\end{array}\right)\int drr
\mathcal{U}_{\kappa'\kappa}^{J(+)}(k,r)
\mathcal{D}_{\kappa_d}(r)\delta_{S'S_d}.
\end{align}
Similarly the reduced $NN$ matrix elements for the magnetic M1
transition are given by
\begin{align}\label{m1}
\mathcal{N}^{\kappa'J}_{\kappa_d\lambda}(k)
=~&\lambda N\frac{(-1)^{L'}}{2M}
\sqrt{3\hat{J}}
\Biggl[(-1)^{S'}\sqrt{L_d(L_d+1)\hat{L}_d}
\left\{
\begin{array}{ccc}
L' & J & S' \\
1 & L_d & 1 \\
\end{array}
\right\}\delta_{S'S_d}
\nonumber\\
&-(-1)^{S_d+J}
\sqrt{6\hat{S}'\hat{S}_d}
\left\{
\begin{array}{ccc}
S' & J & L' \\
1 & S_d & 1 \\
\end{array}\right\}\mu_\pm[(-1)^{S_d}\pm(-1)^{S'}]
\left\{
\begin{array}{ccc}
\frac{1}{2} & S' & \frac{1}{2} \\
S_d & \frac{1}{2} & 1 \\
\end{array}\right\}\Biggr]\times\nonumber\\
&\int dr
\mathcal{U}_{\kappa'\kappa}^{J(+)}(k,r)
\mathcal{D}_{\kappa_d}(r)\delta_{L'L_d}.
\end{align}
with $\mu_+=\mu_s$ and $\mu_-=\mu_v$. The reduced matrix elements
for the magnetic M1 $N\rightarrow\Delta$ transitions of particle 1
(for isospin transitions $0\leftrightarrow1$ and $1\leftrightarrow2$)
\begin{align}\label{md1}
\Delta^{\kappa'J}_{\kappa_d\lambda}(k)
=~&\lambda N\frac{\mu^\star}{M}
(-1)^{L'+J}
\sqrt{
8\hat{J}\hat{S}'\hat{S}_d}
\left\{
\begin{array}{ccc}
S' & J & L' \\
1 & S_d & 1 \\
\end{array}\right\}
\left\{
\begin{array}{ccc}
\xi' & S' & \frac{1}{2} \\
S_d & \xi & 1 \\
\end{array}\right\}\times\nonumber\\
&\int dr
\mathcal{U}_{\kappa'\kappa}^{J(+)}(k,r)
\mathcal{D}_{\kappa_d}(r)\delta_{L'L_d},
\end{align}
where $\xi$ is the degree of freedom for spin: 1/2 for a 
nucleon and 3/2 for a Delta. In the transitions of particle
2 eq. \eqref{md1} has an additional phase of $-(-)^{S_d+S'}$.
Totally the M1 reduced matrix elements are given by
$\mathcal{M}^{\kappa'J}_{\kappa_d\lambda}(k)=
\mathcal{N}^{\kappa'J}_{\kappa_d\lambda}(k)+
\Delta^{\kappa'J}_{\kappa_d\lambda}(k)$.
For convenience we also define
$\mathcal{M}^{\kappa'J}_{\kappa_d\lambda}(k)=
\lambda\mathcal{M}^{\kappa'J}_{\kappa_d}(k)$.
Despite the multiplication factor, the general results 
\eqref{e1} and \eqref{m1} reduce to the expressions of ref. 
\cite{liu} for the five lowest amplitudes. 
%
%
%\bibliography{references}

%
%
\end{document}